\documentclass[showpacs,twocolumn,superscriptaddress]{revtex4}
\usepackage{multirow}
\usepackage{graphicx}
\usepackage{amsmath}

\def\version{version}
\newcommand{\sqrtsNN}{\mbox{$\sqrt{\mathrm{\it s_{NN}}}$} }
\newcommand{\vtwo}{$v_{2}$ }

\newcommand{\ks}{${K}^{0}_{S}$ }

\newcommand{\lam}{$\Lambda$ }

\def \auau  {Au+Au }
\def \cucu  {Cu+Cu }

\def \eparttwos {$\varepsilon_{\mathrm{part}}\{2\}$}

\def \lt {\mbox{$<~$} }
\def \gt {\mbox{$>~$} }

\usepackage{color}

\begin{document}
\title{
\begin{flushright}
{
\small \sl \version \\
}
\end{flushright}
An experimental review on elliptic flow of strange and multi-strange hadrons in relativistic heavy ion collisions}

\author{Shusu Shi}

\address{Key Laboratory of Quarks and Lepton Physics (MOE) and Institute of Particle Physics, Central China Normal University, Wuhan, 430079, China}

\date{\today}

\begin{abstract}
Strange hadrons, especially multi-strange hadrons are good probes for the early partonic stage of heavy ion collisions due to their small hadronic
cross sections. In this paper, I give a brief review on the elliptic flow measurements of strange and multi-strange hadrons in relativistic heavy ion collisions at 
Relativistic Heavy Ion Collider (RHIC) and Large Hadron Collider (LHC).
 
\end{abstract}

\maketitle
\section{Introduction}
\label{sect_intro}

\begin{figure*}[ht]
\vskip 0cm
\begin{center} \includegraphics[width=0.5\textwidth]{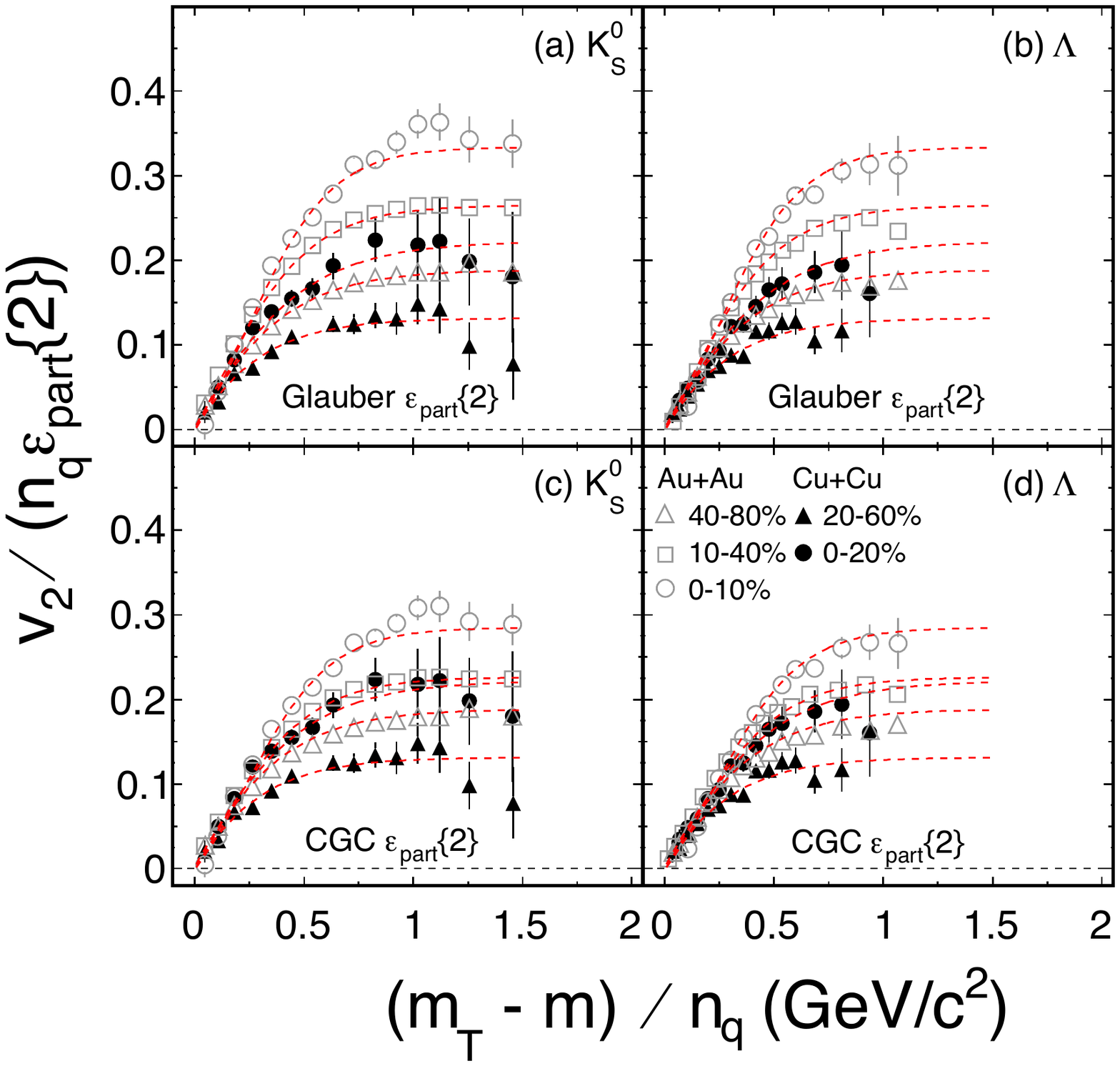}\end{center}
\caption{ Centrality dependence of \vtwo scaled by number of quarks and participant eccentricity
($v_2/$($n_q$\eparttwos)) for \ks and \lam as a function of ($m_T - m$)/$n_q$ in $0-10$\%, $10-40$\% and $40-80$\% \auau collisions
and $0-20$\% and $20-60$\% \cucu collisions at \sqrtsNN = 200 GeV~\cite{cucu}. }
\label{figure1}
\end{figure*}

\begin{figure*}[ht]
\vskip 0cm
\begin{center} \includegraphics[width=0.5\textwidth]{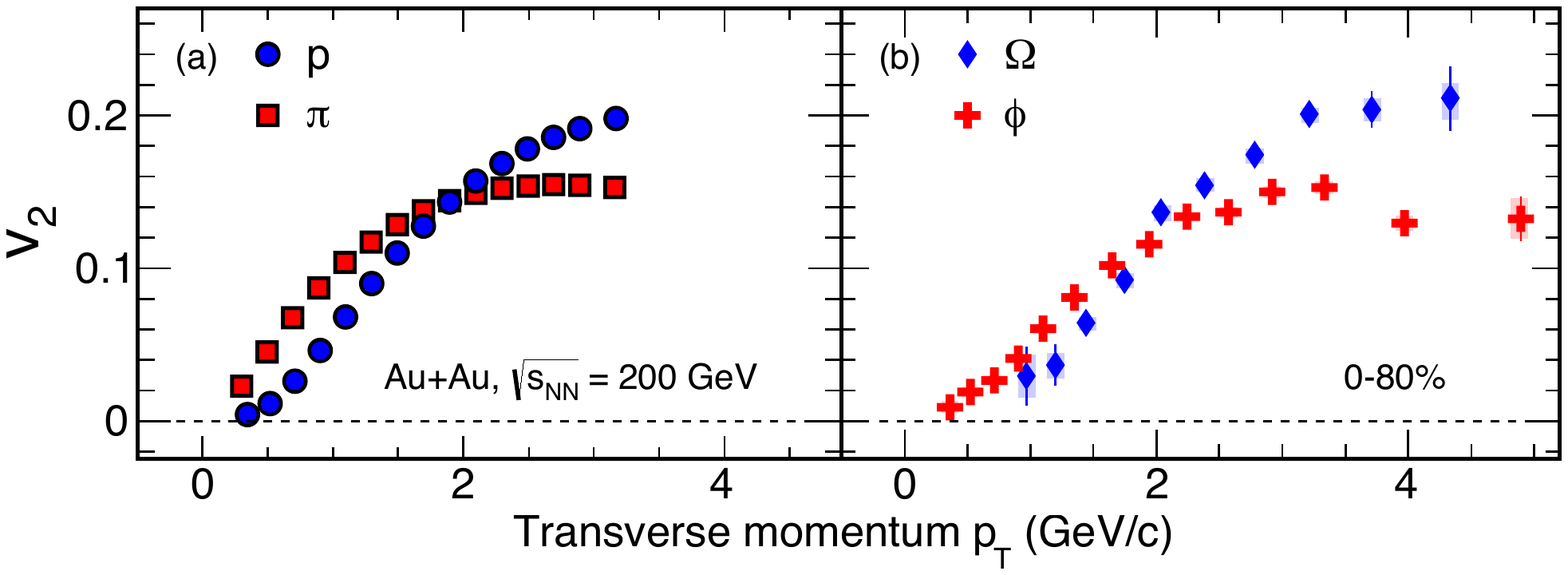}\end{center}
\caption{The $v_{2}$ as function of $p_{T}$ for (a) $\pi$, proton and (b) $\phi$, $\Omega$ in Au+Au 
collisions at $\sqrt{s_{NN}}$ = 200 GeV for $0-80\%$ centrality~\cite{paper}.}
\label{figure2}
\end{figure*}

\begin{figure}[t]
\vskip 0cm
\includegraphics[width=0.5\textwidth]{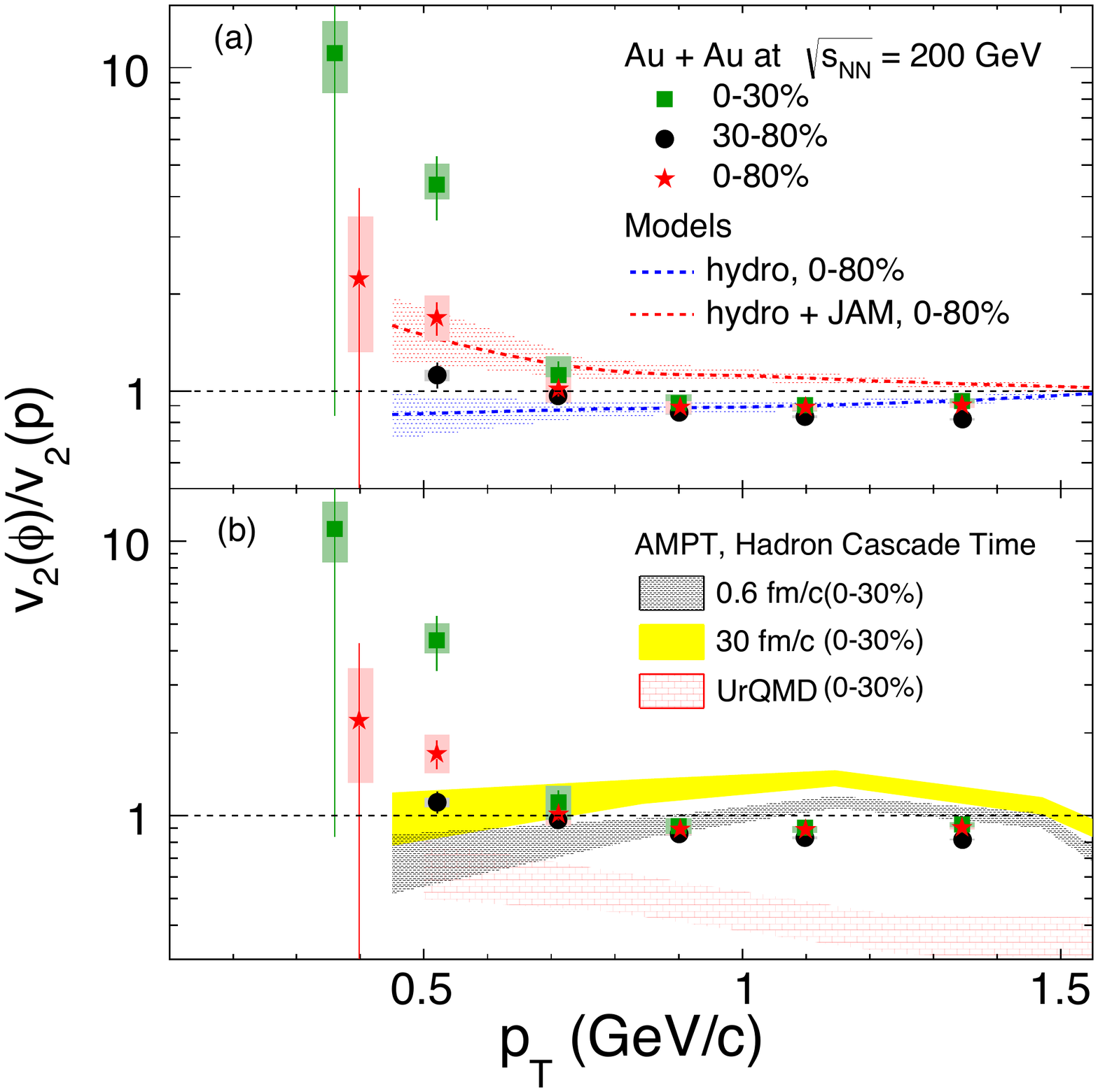}
\caption{The ratio of $v_{2}(\phi)$ to $v_{2}(p)$ as function of $p_{T}$ in Au + Au collisions at
$\sqrt{s_{NN}}$ = 200 GeV for two centralities, $0-30\%$ and $30-80\%$ 
The bands in panel (a) and (b) represent the hydro and transport model calculations for
$v_{2}(\phi)/v_{2}(p)$, respectively~\cite{paper}. }
\label{figure3}
\end{figure}

\begin{figure*}[ht]
\vskip 0cm
\begin{center} \includegraphics[width=0.7\textwidth]{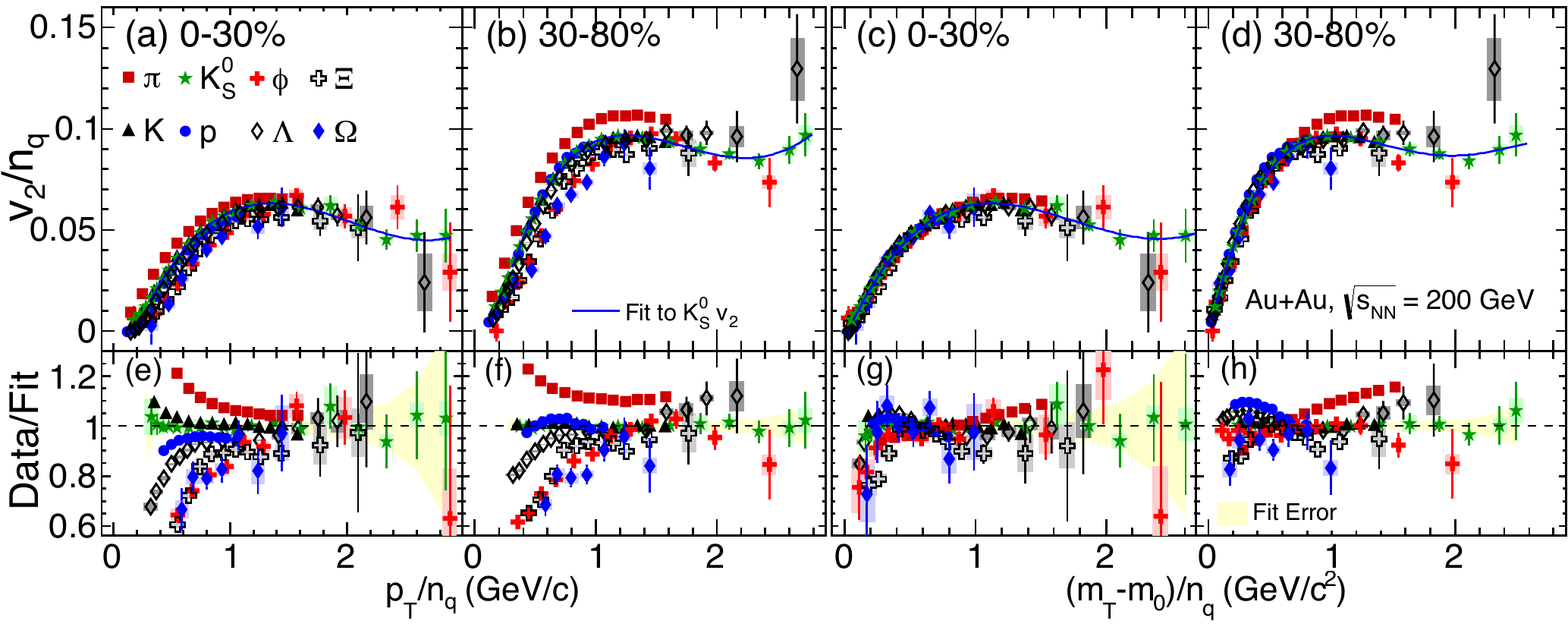}\end{center}
\caption{$v_{2}$ scaled by number of constituent quarks ($n_{q}$)  as a function of $p_{T}/n_{q}$ and 
$(m_{T}-m_{0})/n_{q}$ for identified hadrons from Au + Au collisions at  $\sqrt{s_{NN}}$ = 200 GeV for two centralities, 
$0-30\%$ and $30-80\%$. 
Ratios with respect to a polynomial fit to $K^{0}_{S}$ $v_{2}$ are shown in the corresponding lower panels~\cite{paper}. }
\label{figure4}
\end{figure*}

\begin{figure*}[ht]
\vskip 0cm
\begin{center} \includegraphics[width=0.6\textwidth]{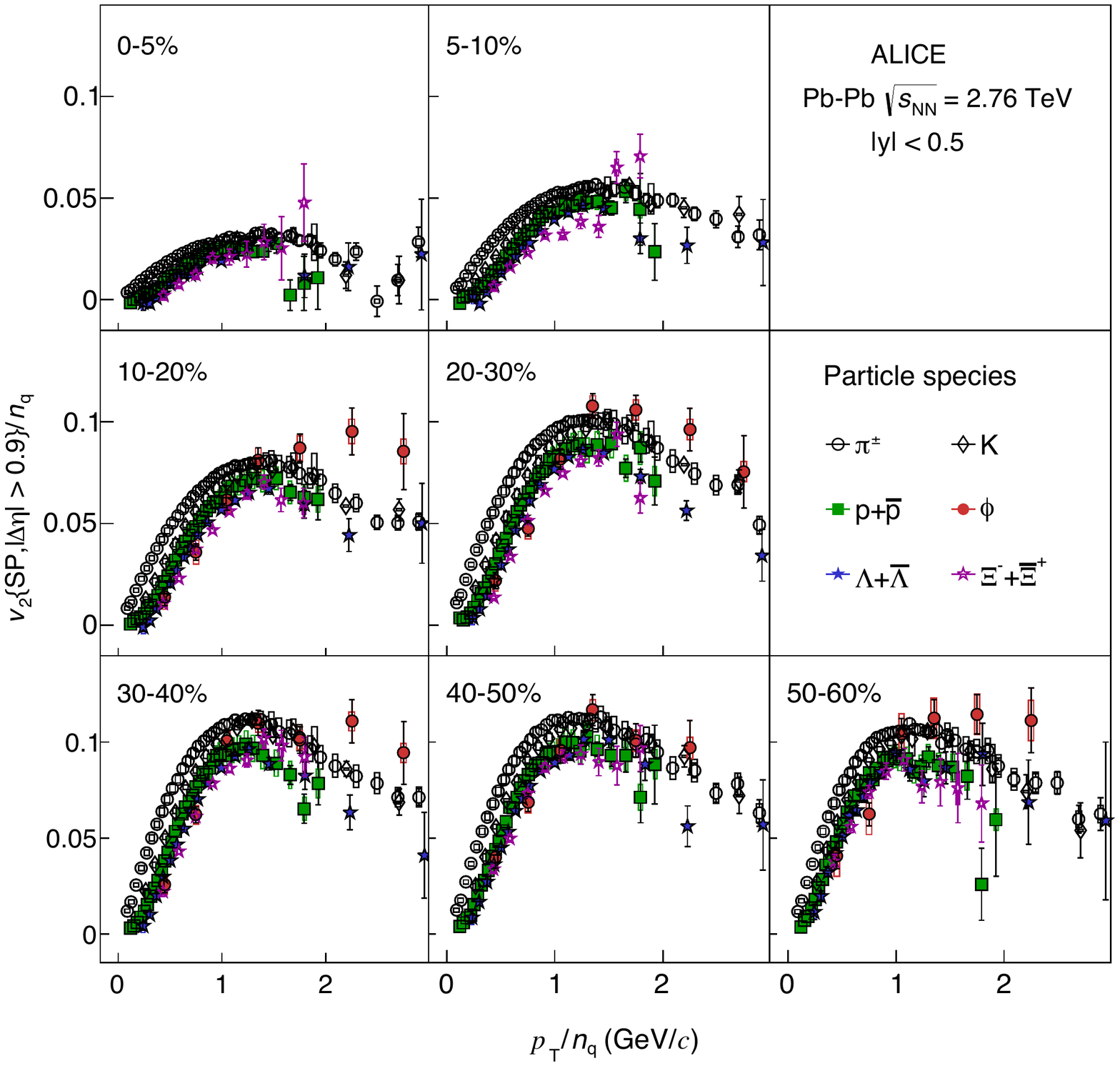}\end{center}
\caption{$v_{2}$ scaled by number of constituent quarks ($n_{q}$)  as a function of $p_{T}/n_{q}$ for
identified hadrons from Pb + Pb collisions at  $\sqrt{s_{NN}}$ = 2.76 TeV for various centrality intervals~\cite{ALICE_flow}.}
\label{figure5}
\end{figure*}

\begin{figure}[t]
\vskip 0cm
\includegraphics[width=0.5\textwidth]{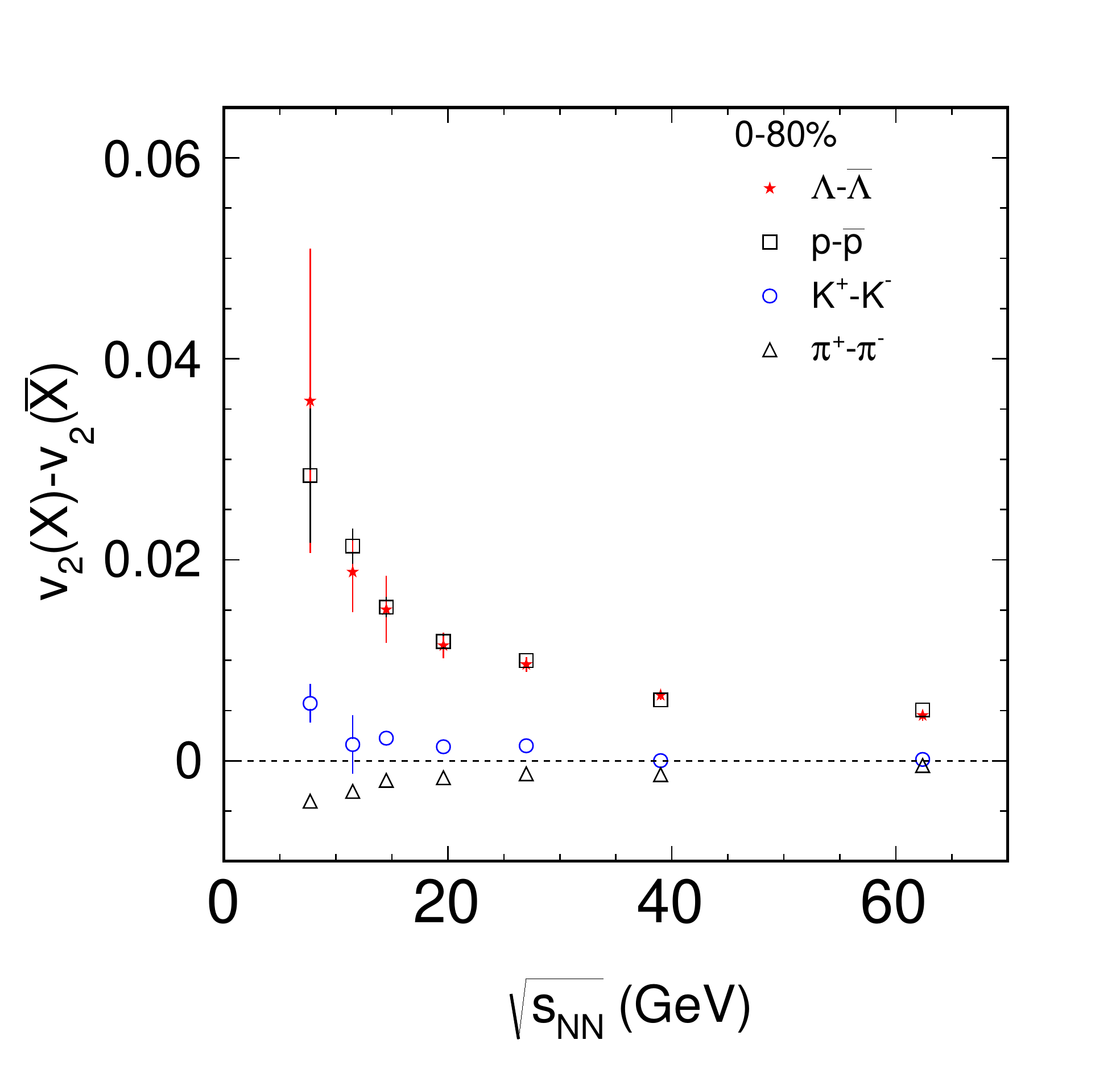}
\caption{The difference in $v_2$ between particles
($X$) and their corresponding antiparticles ($\bar{X}$) as a function of beam energy for $0-80$\% central Au + Au collisions~\cite{BES2, BES3}.}
\label{figure6}
\end{figure}

\begin{figure}[t]
\vskip 0cm
\includegraphics[width=0.5\textwidth]{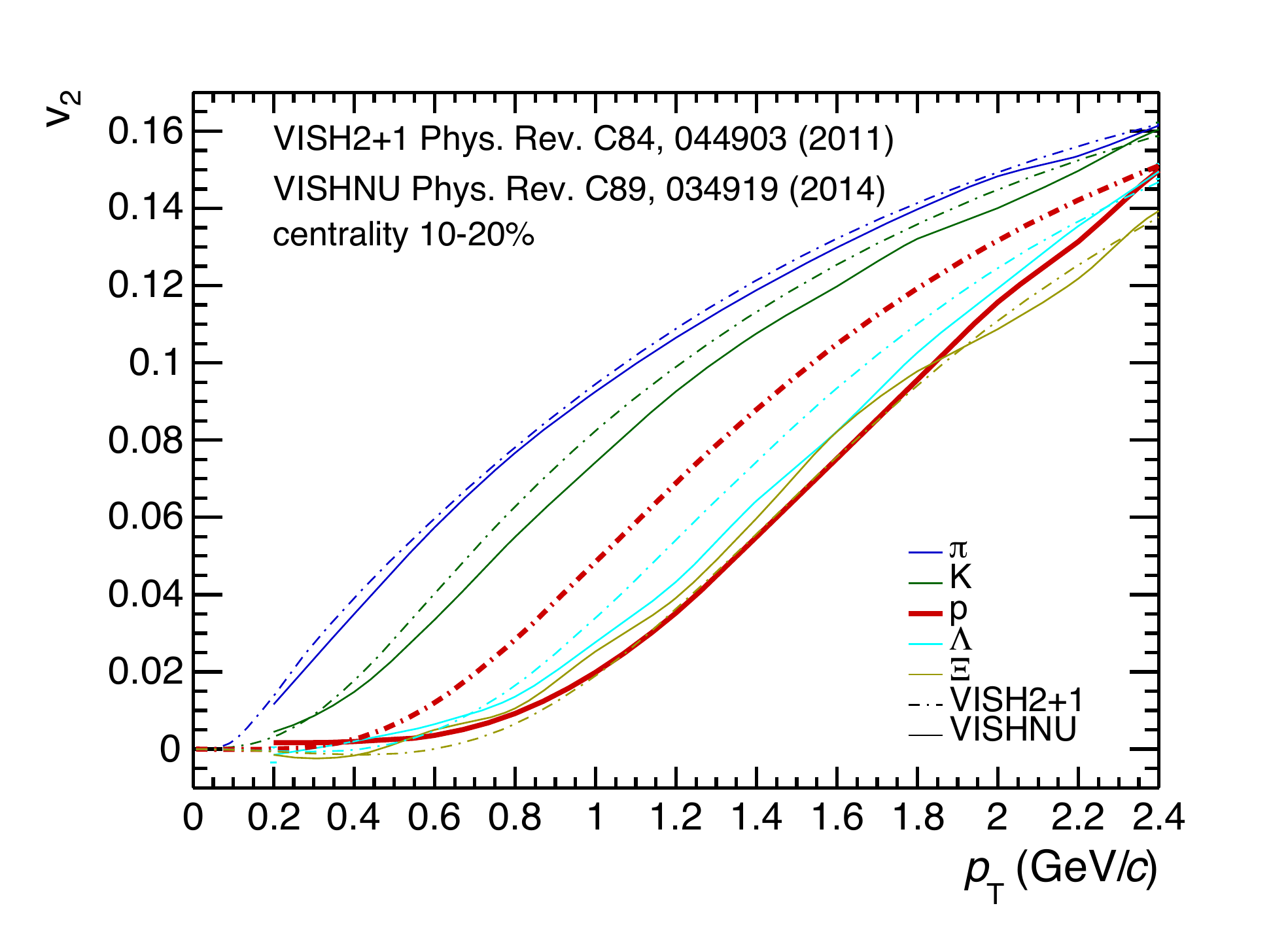}
\caption{The viscous hydrodynamical calculations without a hadronic cascade afterburner (VISH2+1)
and with a hadronic cascade afterburner(VISHNU)~\cite{raimond, Song2011, Song2014, Song2015}  }
\label{figure7}
\end{figure}

\begin{figure}[t]
\vskip 0cm
\includegraphics[width=0.5\textwidth]{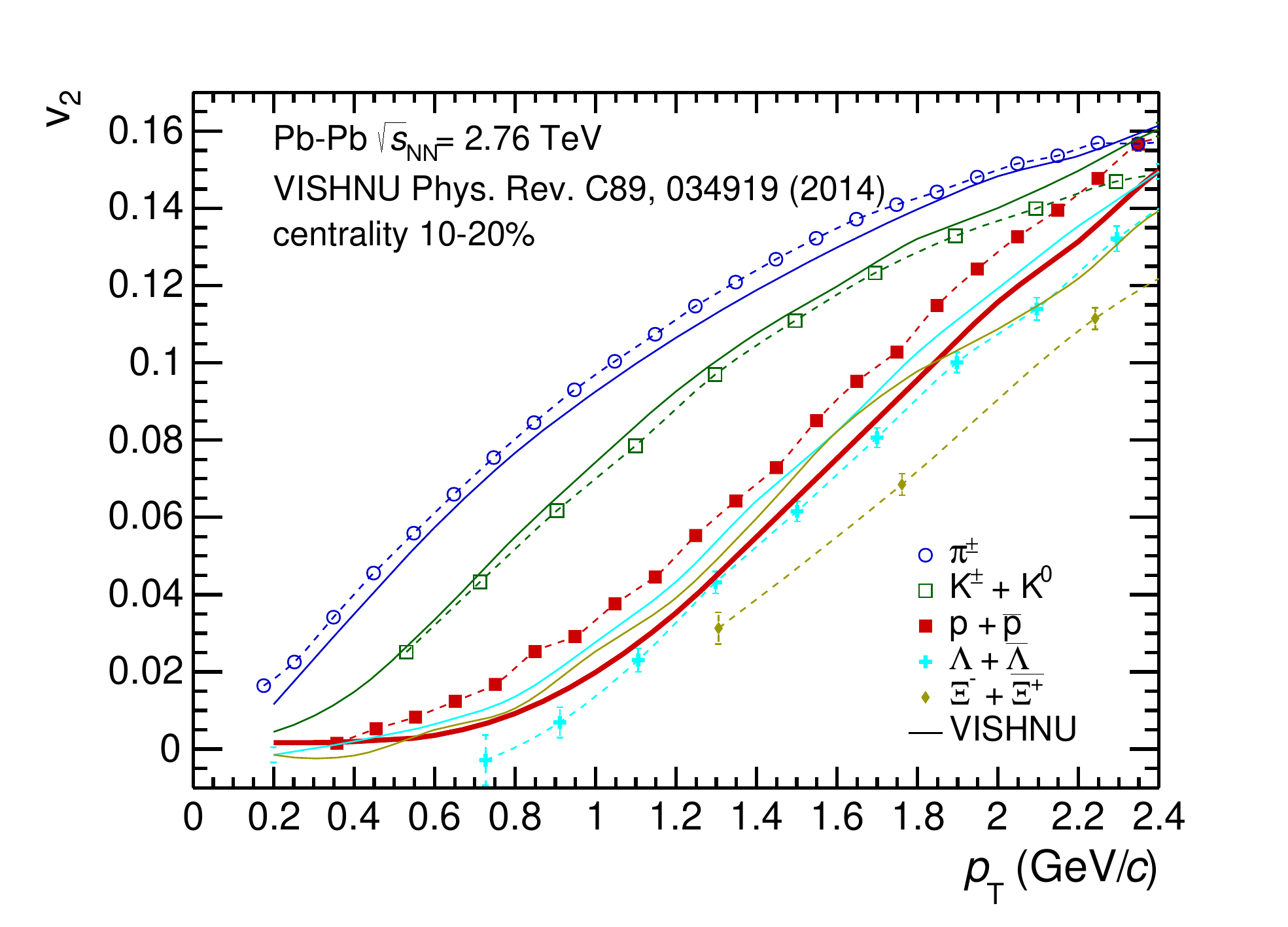}
\caption{The comparison of ALICE measurements and VISHNU model calculations of $v_2$ as a function of $p_T$ at $\sqrt{s_{NN}}$ = 2.76 TeV 
for $10-20$\% central collisions~\cite{ALICE_flow, raimond, Song2011, Song2014, Song2015}.}
\label{figure8}
\end{figure}

At the early stage of high energy relativistic heavy ion collisions, a hot and dense, strongly interacting medium named Quark Gluon Plasma (QGP) is created~\cite{rhicwp1, rhicwp2}. 
The subsequent system evolution is determined by the nature of the medium. 
Experimentally, the dynamics of the system evolution has been studied by measuring the azimuthal anisotropy of the particle production relative to the reaction plane~\cite{flow1, flow2, review}. 
The centrality of the collision, defined by the transverse distance between the centers of the colliding nuclei called the impact parameter, 
results in an `almond-shaped' overlap region that is spatially azimuthal anisotropic. It is generally assumed that the initial spatial anisotropy 
in the system is converted into momentum-space anisotropy through re-scatterings~\cite{rescatterings}.
The elliptic flow, $v_{2}$, which is the second Fourier coefficient of the azimuthal distribution of produced particles with respect
to the reaction plane, is defined as $v_{2}=\langle\cos 2(\varphi-\Psi)\rangle$, where $\varphi$ is the azimuthal angle of produced particle and $\Psi$
is the azimuthal angle of the reaction plane. 
The initial anisotropy in the coordinate space diminishes rapidly as the system expands. Thus, the driving force of $v_2$ quenches itself.
Due to the self-quenching effect, the elliptic flow provides information about the dynamics at the early stage of the collisions~\cite{physics1, physics2, physics3}. Elliptic flow can provide information about the pressure gradients, the effective degrees of freedom, the degree of thermalization, and equation of state of the matter created at the early stage~\cite{review}. 
However, early dynamic information might be obscured by later hadronic rescatterings~\cite{hyrdo_cascade1, hyrdo_cascade2}.
Strange hadrons, especially multi-strange hadrons and the $\phi$ meson are believed to be less sensitive to hadronic rescatterings in the late stage of collisions, as their
freeze-out temperatures are close to the phase transition temperature and
their hadronic interaction cross sections are expected to be small~\cite{multistrange1, multistrange2}.
In this paper, I am going to review the elliptic flow results of strange and multi-strange hadron in relativistic heavy ion collisions from RHIC to LHC energies.

\section{Discussions}

\subsection{Centrality and system size dependence}



The values of $v_2$ are usually divided by the initial spatial anisotropy,
eccentricity, to remove the geometric effect in order to study the centrality and system size dependence of $v_2$.
The participant eccentricity is the initial configuration space eccentricity of the participants which is defined by~\cite{ecc1,cucu}
 \begin{equation}
\varepsilon_{\rm part} = \frac{\sqrt{(\sigma_{y}^{2} - \sigma_{x}^{2})+4(\sigma_{xy}^{2})}}{\sigma_{y}^{2} + \sigma_{x}^{2}},
\label{Equation:partiecc}
\end{equation}
In this formula, $\sigma_{x}^{2} = \langle x^{2} \rangle - \langle x\rangle^{2}$,
$\sigma_{y}^{2} = \langle y^{2} \rangle - \langle y\rangle^{2}$ and
 $\sigma_{xy} = \langle xy \rangle - \langle x \rangle\langle y \rangle$,
with $x$, $y$ being the position of the participating nucleons in the transverse plane.
The root mean square of the participant eccentricity
\begin{equation}
\varepsilon_{\rm part}\{2\} = \sqrt{\langle\varepsilon_{\rm part}^{2}\rangle},
\label{Equation:epart2}
\end{equation}
is calculated from the Monte Carlo Glauber model~\cite{glauber} and Color Glass Condensate (CGC) model~\cite{CGC}.

Figure~\ref{figure1} shows the centrality and system size dependence of \ks and $\Lambda$
$v_2$ in $\sqrt{s_{NN}}$ = 200 GeV heavy ion collisions~\cite{cucu}. The eccentricity scaled $v_2$ has been further normalized by number of constituent quark ($n_q$) to
make \ks and $\Lambda$ results follow a same curve.
The results from $0-20$\% and $20-60$\% central Cu + Cu collisions and from $0-10$\%,
$10-40$\% and $40-80$\% central Au + Au collisions are presented.
For a given collision system, stronger
collectivity flow is apparent as higher scaled $v_{2}$ values in more central collisions. 
For both Au + Au and Cu + Cu collisions, larger collective flow is observed in larger system size 
which could be characterized by number of participants. 
Namely, the collisions with larger number of participants generate larger
collective flow.

\subsection{Multi-strange hadron and $\phi$ meson $v_2$}

STAR experiment presented the first $v_2$ results of multi-strange hadrons based on $2 \times 10^{6}$ events collected in the year of 2001 - 2002~\cite{STAR_multi-strange}.
Significant $v_2$ signals of $\Xi$ baryons which are similar to results for $\Lambda$ baryons are observed in Au + Au collisions at $\sqrt{s_{NN}}$ = 200 GeV.
At low $p_T$ ($\lt$ 2 GeV/$c$), the mass ordering is observed for $\Xi$ $v_2$ which is in agreement with the hydrodynamic model calculations.
Due to limited statistics, the $v_2$ of $\Omega$ baryons have large statistical uncertainties, it is not clear whether $\Omega$ $v_2$ follows baryon or meson band
at the intermediate $p_T$ range(2 - 5 GeV/$c$). But non-zero value of $v_2$ was clearly observed at that time. These results suggest that collective motion has been developed
at parton phase in Au + Au collisions at $\sqrt{s_{NN}}$ = 200 GeV.

Later, in the RHIC runs of the year 2010-2011, about 730 million minimum bias events were recorded by STAR.  
Sufficient statistics of multi-strange hadrons and $\phi$ mesons support the precise measurements on $v_2$.
The multi-strange hadrons and the $\phi$ meson 
were reconstructed though the following decay channels:
$\phi$ $\rightarrow$ $\it{K}^{+}$ + $\it{K}^{-}$,
$\Xi^{-}$ $\rightarrow$ $\Lambda$ + $\pi^{-}$ ($\overline{\Xi}^{+}$ $\rightarrow$ $\overline{\Lambda}$ + $\pi^{+}$) and
$\Omega^{-}$ $\rightarrow$ $\Lambda$ + $\it{K}^{-}$ ($\overline{\Omega}^{+}$$\rightarrow$ $\overline{\Lambda}$ + $\it{K}^{+}$).
Figure~\ref{figure2} shows the $v_{2}$ as a function of $p_{T}$ for (a) $\pi$, protons  and (b) $\phi$, $\Omega$  in Au + Au collisions at
$\sqrt{s_{NN}}$ = 200 GeV for $0-80\%$ centrality~\cite{paper, paper1}. 
A comparison between $v_{2}$ of $\pi$ and protons, consisting of up ($\it u$) and down ($\it d$) light constituent quarks is shown in panel (a). 
Correspondingly, panel (b) shows a comparison of $v_{2}$ of $\phi$ and $\Omega$ containing $\it s$ constituent quarks. 
This is the first time that high precision measurement of $\Omega$ baryon $v_2$ up to 4.5 GeV/$c$ is available in experiments of heavy ion collisions.
In the low $p_{T}$ region ( $p_{T}$ $\lt$ 2.0 GeV/$c$), the $v_{2}$ of  $\phi$ and $\Omega$ follows mass ordering.
At intermediate $p_{T}$  ( 2.0 $\lt$ $p_{T}$ $\lt$ 5.0 GeV/$c$), a baryon-meson separation is observed.
The $v_2$ results of $\phi$ mesons are consistent in two independent measurements at RHIC, PHENIX~\cite{PHENIX_phi} and STAR.
It is evident that the $v_{2}(p_{T}) $ of hadrons consisting only of strange constituent quarks ($\phi$ and $\Omega$)  is similar to that of light hadrons, $\pi$ and protons. 
However the $\phi$ and $\Omega$ do not participate strongly in the hadronic interactions, because of the smaller hadronic cross sections compared to $\pi$ and protons. 
It suggests the major part of the collectivity is developed during the partonic phase in high energy heavy ion collisions.
ALICE experiment recently published multi-strange hadron and $\phi$ meson $v_2$ measurements in Pb + Pb collisions at $\sqrt{s_{NN}}$ = 2.76 TeV~\cite{ALICE_flow}.  
Also significant $v_2$ values for these particles are observed.
Experimental measurements at RHIC and LHC indicates partonic collectivity has been built up in high energy heavy ion collisions.

\subsection{Comparison of $\phi$ meson and proton $v_2$}

The $\phi$ meson and proton show different sensitivity on the hadronic rescatterings. As discussed previously,
the $\phi$ meson is less sensitive to the late hadron hadron interactions than light hadrons due to the smaller hadronic cross section.
It means light hadrons (e.g. protons) would gain larger additional radial flow which modifies the $v_2$($p_T$) shape during final hadronic rescatterings.
Hydrodynamical model calculations predict that $v_{2}$ as a function of $p_{T}$ for different particle species follows mass ordering, 
where the $v_{2}$ of heavier hadrons is lower than that of lighter hadrons~\cite{hydro}. The identified hadron $v_{2}$ measured in experiment indeed proves the mass ordering in the
low $p_{T}$ region ($p_{T}$ $\lt$ 2.0 GeV/$c$).
Hirano {\it et al.} predict the mass ordering of $v_{2}$ could be broken between $\phi$ mesons and protons at low $p_{ T}$ ($p_{T}$ $\lt$ 1.5 GeV/$c$)
based on a model with ideal hydrodynamics plus hadron cascade process~\cite{hyrdo_cascade1, hyrdo_cascade2}.
Here $\phi$ mesons and protons are chosen for the study, as their rest masses are quite close to each other.
As the model calculations assign a smaller hadronic cross section for $\phi$ mesons compared to protons, 
the broken mass ordering is regarded as the different hadronic rescattering contributions on the $\phi$ meson and
proton $v_2$.

Figure~\ref{figure3} shows the ratios of $\phi$ $v_{2}$ to proton $v_{2}$ from model calculations and experimental data~\cite{paper, paper1}. 
This ratio is larger than unity at $p_{T}$ $\sim$ 0.5 GeV/$c$ for 0-30$\%$ centrality. It indicates
breakdown of the expected mass ordering in that momentum range. This could be due to a large effect of hadronic
rescatterings on the proton $v_{2}$. The data of 0-80\% centrality around 0.5 GeV/$c$ is quantitatively agrees with hydro + hadron cascade calculations indicated 
by the shaded red band in panel (a) of Fig.~\ref{figure3}, even though there is a deviation in higher $p_T$ bins. 
A centrality dependence of $v_{2}(\phi)$ to $v_{2}(p)$ ratio is observed in the experimental data. Namely, 
the breakdown of mass ordering of $v_{2}$ is more pronounced in 0-30$\%$
central collisions than in 30-80$\%$ peripheral collisions. 
In the central events,  both hadronic and partonic interactions are stronger than in peripheral events.
Therefore,  the larger effect of late stage hadronic interactions relative to the partonic collectivity
produces a greater breakdown of mass ordering in the 0-30$\%$ centrality data than in the 30-80$\%$.
This observation indirectly supports the idea that the $\phi$ meson has a smaller hadronic interaction cross section.  
The ratio of $\phi$ $v_{2}$ to proton $v_{2}$ was also studied by using the transport models AMPT~\cite{ampt} and UrQMD~\cite{urqmd}. 
The panel (b) of Fig.~\ref{figure3} shows the $v_{2}(\phi)$ to $v_{2}(p)$ ratio for 0-30$\%$ centrality from AMPT and UrQMD models.
The black shaded band is from AMPT with a hadronic cascade time of 0.6 fm/$c$ while the yellow
band is for a hadronic cascade time of 30 fm/$c$.  Larger hadronic cascade time is equivalent to stronger hadronic interactions.
It is clear that the $v_{2}(\phi)/v_{2}(p)$ ratio increases with increasing hadronic cascade time. 
This is attributed to a decrease in the proton $v_{2}$ due to an  increase in hadronic re-scattering
while the $\phi$ meson $v_{2}$ is less affected. The ratios from the UrQMD model
are much smaller than unity (shown as a brown shaded band in the panel (b) of Fig.~\ref{figure3}).
The UrQMD model lacks partonic collectivity, thus the $\phi$ meson $v_{2}$ is not fully developed.
Non of these models could describe the detailed shape of the $p_T$ dependence.
In $\sqrt{s_{NN}}$ = 2.76 TeV Pb + Pb collisions at LHC, there is an indication that the $\phi$ meson $v_2$ is 
larger than the proton $v_2$ for the lowest $p_T$ bin~\cite{ALICE_flow, raimond}. 
Unfortunately, currently the uncertainties on the ALICE $\phi$ meson $v_2$ measurements are too large to conclude.

\subsection{Number of constituent quark scaling}

The Number of Constituent Quark (NCQ) scaling in $v_{2}$ in the intermediate $p_{T}$ range
(2 $\lt$ $p_T$ $\lt$ 5 GeV/$c$) could be well reproduced by the
quark coalescence~\cite{cola} or recombination~\cite{recom} mechanisms in
particle production. The NCQ scaling indicates the collectivity in the parton level has been
achieved in high energy heavy ion collisions at RHIC. 
Figure~\ref{figure4} shows number of constituent quarks ($n_{q}$) scaled $v_{2}$ as a function of transverse momentum scaled by 
$n_{q}$ ($p_{T}/n_{q}$) and transverse mass minus rest mass scaled by $n_{q}$ ($(m_{T}-m_{0})/n_{q}$) for
identified hadrons from Au + Au collisions at  $\sqrt{s_{NN}}$ = 200 GeV 
for two centralities, 0-30$\%$ and 30-80$\%$. 
To investigate the possible system size dependence deviation from NCQ scaling, 
the $K^{0}_{S}$ $v_{2}$ was fitted with a third-order polynomial function.  
Then the ratio to the $K^{0}_{S}$ fit was calculated.  The lower panels of Fig.~\ref{figure4} show the results.
Excluding pions, the scaling holds approximately within 10$\%$ for both 0-30$\%$ and 30-80$\%$ centralities. 
The pion is excluded as it is strongly affected by resonance decay process and non-flow correlations~\cite{dongx}. 
Figure~\ref{figure5} shows NCQ scaling at LHC energy.
The maximum deviation from  NCQ scaling is $\sim$20$\%$ at $\sqrt{s_{NN}}$ =2.76 TeV as observed by ALICE experiment~\cite{ALICE_flow}. 
Therefore, at top RHIC energy, NCQ scaling holds better than LHC energy.
 
Recently, CMS collaboration presented the $v_2$ results of strange hadrons ($K_{S}^{0}$ and $\Lambda$) 
in $p$ + Pb collisions at $\sqrt{s_{NN}}$ = 5.02 TeV with
event sample of large multiplicity~\cite{CMS1, CMS2}. 
A nice NCQ scaling (less than 10\% violation) is observed. It indicates the partonic level collectivity has
been built up even in small $p$ + Pb colliding system. It would be interesting to compare the NCQ scaling 
using event samples with large and small multiplicity in the future.

\subsection{Beam energy dependence}

STAR experiment has covered the beam energies of $\sqrt{s_{NN}}$ = 7.7, 11.5, 14.5, 19.6, 27, 39, 62.4 and 200 GeV. 
During 2010 - 2014, a Beam Energy Scan program (phase I) was carried out at RHIC. 
The main motivation is to explore the nuclear matter phase structure in the higher 
net-baryon region. 

The most striking feature on the $v_2$ measurements is the observation of an energy dependent difference in $v_2$
between particles and their corresponding antiparticles~\cite{{BES1, BES2}}.
Figure~\ref{figure6} shows the difference in $v_2$ between particles and their corresponding antiparticles as a function of 
beam energy. The difference between baryon and anti-baryon is much more pronounced than difference between mesons.
Proton versus anti-proton and $\Lambda$ versus $\bar{\Lambda}$ show same magnitude of difference.
This difference naturally breaks the number of constituent quark scaling (NCQ) in $v_2$ which is 
regarded as an evidence of partonic collectivity in the top energy heavy ion collisions at RHIC. It indicates the hadronic degrees of
freedom play a more important role at lower collision energies.
The data have also been compared to hydrodynamics + transport (UrQMD) hybrid model~\cite{hybrid} and Nambu-Jona-Lasino (NJL) model~\cite{NJL} which considers both partonic and hadronic potential. The hybrid model could reproduce the baryon (proton) data, but fails to explain the mesons; whereas the NJL model could qualitatively reproduce the hadron splitting.  However, even if one tunes the $R_v$ parameter which is related to the partonic potential, NJL model fails to reproduce the magnitude for all hadron species simultaneously. 
Analytical hydrodynamic solution can reproduce the data within uncertainties~\cite{hydro_solution}. It predicts $\Delta v_{2}^{p} \gt \Delta v_{2}^{\Lambda} \gt \Delta v_{2}^{\Xi} \gt \Delta v_{2}^{\Omega}$
for baryons. Future high precise data will clarify the validity of this description. 

\subsection{Comparison with hydrodynamic calculations}

The $p_T$ differential $v_2$ could be modified by an increase on both collective and radial flow with increasing of 
colliding energy. It is qualitatively described by hydrodynamic calculations~\cite{Song2011}. 
The recent comparison between ALICE measurements and model calculations
shows a nice agreement in $40-50$\% central collisions including strange baryon $\Lambda$ and multi-strange baryon $\Xi$.
However, for more central collisions (e.g. $10-20$\%) a clear discrepancy is 
observed for protons, $\Lambda$ and $\Xi$~\cite{ALICE_flow}.

Later, it was realized the hadronic rescatterings is important to be included in the hydrodynamic calculations for a fair comparison between
data and models~\cite{Song2014}.
In Fig.~\ref{figure7}, viscous hydrodynamical calculations with(VISHNU)  and without (VISH2+1)  a hadronic cascade afterburner
are compared. 
The increase in mass splitting between identified particles for {\sc VISHNU} (solid curves) compared to 
{\sc VISH2+1} (dashed curves) illustrates the larger radial flow in the {\sc VISHNU} calculations due to the contribution 
of the hadronic cascade. 
The mass splitting between the pions and strange baryons ($\Lambda$) / multi-strange baryons ($\Xi$) does not change much,
as small hadronic rescattering cross sections are assigned to these particles.  
The mass ordering observed in pure viscous hydrodynamical calculations is not preserved anymore
between protons and strange baryons ($\Lambda$) / multi-strange baryons ($\Xi$)  after including
the hadronic interactions in VISHNU.
Figure~\ref{figure8} shows the comparison between the $p_T$-differential $v_2$ measured by ALICE and the {\sc VISHNU} model. 
Even though VISHNU gives a very well description of kaons; clear discrepancy for protons, $\Lambda$ and $\Xi$ is observed. 
The VISHNU calculations under-predict the $v_2$ of protons and over-predict the $v_2$ of $\Lambda$ and $\Xi$.
Obviously, the current theoretical framework of viscous hydrodynamics plus a hadron cascade afterburner does not  
describe the $v_2$ as a function of $p_T$ for identified particles in more central collisions better.
One of the possible reasons is that hadronic interaction process for some particle species might not be well understood.

\section{Summary}
\label{sect_summary}

In this paper, I review the elliptic flow results of strange and multi-strange hadrons in relativistic heavy ion collision from RHIC to
LHC energies. The centrality and system size dependence of $v_2$ could be described by 
number of participants in both \auau and \cucu collisions at $\sqrt{s_{NN}}$ = 200 GeV.
The precise measurements of multi-strange hadron $v_2$, especially for the $\Omega$ baryons indicates the collectivity
has been built-up in the early partonic stage of collisions. 
The comparison between the $v_2$ of $\phi$ mesons and protons shows a possible violation of hydrodynamics inspired mass ordering in
$0-30$\% central collisions. It can be qualitatively explained by the different effects of late hadronic interactions on the $\phi$ meson and proton $v_2$.
The NCQ scaling of identified particles in top energy heavy ion collisions at RHIC is better than LHC energy suggesting that coalescence might be the dominant hadronization mechanism at RHIC in the intermediate transverse momentum region (2 $\lt$ $p_T$ $\lt$ 5 GeV/$c$). Also, the NCQ scaling is observed in small colliding system, $p$ + Pb, at $\sqrt{s_{NN}}$ = 5.02 TeV.
It indicates the partonic level of collectivity has also been reached in high energy $p$ + Pb collisions. At lower beam energy ($<$ $\sqrt{s_{NN}}$ = 39 GeV), 
a difference is observed between $v_2$ values of particles and anti-particles. Currently there is no theoretical framework can reproduce the data quantitatively.
The recent comparison between viscous hydrodynamic calculations with a hadronic cascade afterburner and experimental data shows a discrepancy on the baryons
which challenges the current knowledge on the hadronic interactions.

\section{Acknowledgments}
This work was supported in part by 
National Basic Research Program of China (973 program) under grand No. 2015CB8569,
the National Natural Science Foundation of China under grant No. 11475070 and 
self-determined research funds of CCNU from the colleges' basic research and operation of MOE under grand No. CCNU15A02039.

%

%
\end{document}